\documentclass[11pt]{article}

\usepackage{amsmath,amsfonts,amssymb,amsthm,bbm,stmaryrd}
\usepackage[dvips]{graphicx}
\usepackage{latexsym}
\usepackage{psfrag}
\usepackage[latin1]{inputenc}
\usepackage{hyperref}
\usepackage{color}
\usepackage{enumitem}
\numberwithin{equation}{section}
\usepackage{float}
\usepackage{multirow}

\newcommand{\be}{\begin{equation}}
\newcommand{\ee}{\end{equation}}
\newcommand{\barray}{\begin{array}}
\newcommand{\earray}{\end{array}}
\newcommand{\bea}{\begin{eqnarray}}
\newcommand{\eea}{\end{eqnarray}}
\newcommand{\bs}{\begin{subequations}}
\newcommand{\es}{\end{subequations}}

\newcommand{\bit}{\begin{itemize}}
\newcommand{\eit}{\end{itemize}}
\newcommand{\bd}{\begin{description}}
\newcommand{\ed}{\end{description}}

\newcommand{\re}{\mathrm{Re}}
\newcommand{\im}{\mathrm{Im}}

\def\la{\langle}
\def\ra{\rangle}

\newcommand{\p}{\partial}

\newcommand{\f}{\frac}

\renewcommand{\d}{\delta}  \newcommand{\eps}{\epsilon} 
 \renewcommand{\th}{\theta}  \newcommand{\vth}{\vartheta} 
    
       \let\r=\rho 
 \newcommand{\s}{\sigma}      
     
\let\Si=\Sigma \let\Om=\Omega

\def\cS{{\cal S}}

\def\cN{{\cal N}}
\def\cR{{\cal R}}

\newcommand{\os}[1]{\overset{\circ}{#1}}

\usepackage{slashed}

\newcommand{\eqonS}{\,\smash{\stackrel{\sscr{\scri}}=}\,}
\newcommand{\eqons}{\,\hat{=}\,}

\usepackage{pgf}
\makeatletter
\newcommand*{\pgfunderleftarrow}{%
  \@ifstar
    {\let\ifpgf@depth\iftrue\mathpalette\@pgfunderleftarrow}
    {\let\ifpgf@depth\iffalse\mathpalette\@pgfunderleftarrow}%
}
\newcommand*{\@pgfunderleftarrow}[2]{%
  #2%
  \edef\pgf@math@fam{\the\fam}%
  \pgfpicture
    \pgfsetbaseline{0pt}
    \pgf@relevantforpicturesizefalse      
    \pgfsetroundcap                       
    \pgfsetarrowsend{to}
    \pgfutil@tempdima=0.28pt%
    \advance\pgfutil@tempdima by.8\pgflinewidth%
    \pgfutil@tempdima-4\pgfutil@tempdima
    \sbox\pgfutil@tempboxa{$\m@th\fam\pgf@math@fam#1#2$}%
    \advance\pgfutil@tempdima-\dp\pgfutil@tempboxa
    \pgfutil@tempdimb\wd\pgfutil@tempboxa
    \pgfpathmoveto{\pgfqpoint{0pt}{\pgfutil@tempdima}}%
    \pgfpathlineto{\pgfqpoint{-\pgfutil@tempdimb}{\pgfutil@tempdima}}%
    \pgfusepath{stroke}
    \ifpgf@depth
      \pgf@relevantforpicturesizetrue
      \pgfpathmoveto{\pgfqpoint{0pt}{-\pgfutil@tempdimb}}%
      \pgfusepath{use as bounding box}%
    \fi
  \endpgfpicture
}
\makeatother

\newcommand{\sscr}{\scriptscriptstyle\rm}

\usepackage{manfnt}
\reversemarginpar

\DeclareFontFamily{U}{matha}{\hyphenchar\font45}
\DeclareFontShape{U}{matha}{m}{n}{
      <5> <6> <7> <8> <9> <10> gen * matha
      <10.95> matha10 <12> <14.4> <17.28> <20.74> <24.88> matha12
      }{}
\DeclareSymbolFont{matha}{U}{matha}{m}{n}
\DeclareMathSymbol{\oright}       {2}{matha}{"69}

\usepackage{pifont}

\usepackage{mathrsfs}
\newcommand{\Dd}{{\mathscr{D}}}
\newcommand{\scri}{{\mathscr{I}}}

\newcommand{\sv}{{\smash{\os{\s}}}}

\newcommand{\DDr}{{\mathbbm D}_\r}

\begin{document}

\title{\bf Spatially local energy density of gravitational waves}

\author{\Large{Antoine Rignon-Bret and Simone Speziale}
\smallskip \\ 
\small{\it{Aix Marseille Univ., Univ. de Toulon, CNRS, CPT, UMR 7332, 13288 Marseille, France}} }
\date{January 6, 2025}

\maketitle

\begin{abstract}
\noindent We propose a new set of BMS charges at null infinity, characterized by a super-translation flux that contains only the `hard' term. 
This is achieved with a specific corner improvement of the symplectic 2-form, and we spell the conditions under which it is unique. 
The charges are associated to a Wald-Zoupas symplectic potential, and satisfy all standard criteria: they are covariant, provide a center-less realization of the symmetry algebra, have vanishing flux in non-radiative spacetimes, and vanish in Minkowski. We use them to define a notion of spatially localized energy density of gravitational waves, and explain how it can be measured doing experiments which are purely local in space and over an extended period of time.
\end{abstract}

\tableofcontents

\section{Introduction}

Bondi's energy loss formula is a cornerstone of our physical
understanding of general relativity. It was the historical theoretical proof that gravitational waves dissipate energy, ending any lingering doubt on their physical existence.
The proof of dissipation refers to the \emph{total} gravitational energy; there is no equivalent statement for a local notion of energy, in line with standard intuition that general covariance prevents the existence of a local energy-momentum tensor for the gravitational field. In this paper we report that general covariance allows an additional step: it is possible to provide a formula for a \emph{spatially local} energy density of gravitational waves at future null infinity which is necessarily dissipated by physical processes.
The only non-locality remaining is over time, in the sense that an extended period of time is needed in order to determine the new energy density from the spacetime curvature. 

Bondi's formula can be understood in the framework of asymptotic symmetries.
The group of gravitational asymptotic symmetries at null infinity has been known since the seminal work of Bondi, van der Burg, Metzner and Sachs (BMS) \cite{Bondi:1962px,Sachs:1962zza} and extensively studied since. It is a generalization of the Poincar\'e group to angle-dependent translations, known as super-translations.
To these asymptotic symmetries are associated asymptotic charges which are conserved in the absence of radiation, and satisfy flux-balance laws in the presence of radiation. While there is a certain amount of ambiguity in the definition of Noether charges and their fluxes, a unique set satisfying certain physical requirements has been identified as early as 
\cite{Geroch:1977jn,Ashtekar:1981hw,Ashtekar:1981bq,Dray:1984rfa,Dray:1984gz}, 
and later related to Noether charges and canonical generators in \cite{Wald:1999wa}, see also discussion in \cite{Ashtekar:2024bpi}.
Bondi's energy loss formula fits elegantly this framework as the flux-balance law for super-translations. It has two contributions, a `hard' term which is monotonic, and a `soft' term which is not. The soft term  vanishes for global translations, and this is crucial to establish Bondi's proof of dissipation for the total energy. The soft term contribution for local super-translations makes their flux non-monotonic, hence preventing an interpretation of these charges in terms of local energy. 
Our main result is the construction of an alternative set of charges whose flux is purely hard for all super-translations.

The uniqueness of the BMS charges relies on a technical assumption concerning the choice of symplectic structure used to define the canonical generators. There is a growing body of evidence that this assumption can be relaxed, allowing for so called `corner improvements', which are important to capture the physics of gauge and gravitational theories in the presence of boundaries, see e.g. \cite{Compere:2008us,Harlow:2019yfa,Freidel:2020xyx,Campiglia:2020qvc,Odak:2021axr,Chandrasekaran:2021vyu,Odak:2022ndm,Donnay:2022hkf,Rignon-Bret:2024wlu}. This is the first technical ingredient to our result. The second is the use of the `covariant shear' introduced in \cite{Compere:2018ylh}, which is related to previous work done in \cite{Ashtekar:1981hw} and to the notion of super-translation  Goldstone field of \cite{Strominger:2013jfa}, which is a corner degree of freedom. With these two inputs, we are able to identify a unique corner-improved symplectic structure whose canonical generators satisfy all the properties of the standard BMS charges, plus the new property of having a purely hard flux. The new super-momentum charge associated with this flux leads to the notion of energy density that we propose.

To motivate the physical interpretation of the new charge as a local energy, we show that 
on top of being a canonical generator for time translations at each point of the celestial sphere, and having a monotonic flux, it is always positive under standard boundary conditions, and extensive. We further describe how it can be measured by an observer localized on the celestial sphere using only the spacetime curvature at that point, over a period of time.

We should emphasize that our proposal is \emph{not} that the new BMS charges presented here should replace the standard ones of \cite{Geroch:1977jn,Ashtekar:1981hw,Ashtekar:1981bq,Dray:1984rfa,Dray:1984gz}. But rather, that both sets are useful in their own right, and better suited for different physical processes. Specifically, if we want to study the memory effect, the standard BMS charges are better suited since their flux contains the soft component. If we want to ask what is the energy dissipated in an arbitrarily small region of the celestial sphere, our new charges are more relevant since they do not have any soft contribution and have a monotonic flux for any future pointing supertranslation. Therefore, depending on the experiment that we aim to perform, we should choose the appropriate set of charges to describe it.
In other words, choosing different charges at $\scri$ is akin to going from Dirichlet to Neumann boundary conditions in conservative systems, or from entropy to free energy in thermodynamics. To be physically relevant, the freedom should be allowed only among charges which satisfy the basic physical criteria of
covariance, conservation in any non-radiative spacetimes, and vanishing in Minkowski spacetime.
These criteria are guaranteed if the charges are derived using the Wald-Zoupas prescription, and this is what we do here.

\section{BMS flux-balance laws}

We assume that the reader has a certain familiarity with null infinity, and refer to the reviews in \cite{Ashtekar:2014zsa,Flanagan:2015pxa,Grant:2021sxk,Ashtekar:2024bpi} for the necessary background. We use the conventions of \cite{Rignon-Bret:2024gcx}, to which we refer for further details.
The Ashtekar-Streubel (AS) flux for a BMS symmetry $\xi$ between two cross-sections $S_1$ and $S_2$ of $\scri$ is given by \cite{Ashtekar:1981bq}
\be \label{ASflux}
	 F_\xi^{\sscr AS} =- \frac{1}{16 \pi} \int_{S_1}^{S_2} N_{ab} \d_\xi \sigma^{ab} \eps_\scri.
\ee   
Here $\s_{ab}$ is the asymptotic shear, which can be assumed without loss of generality to refer to an affine foliation of $\scri$, and then the news tensor is $N_{ab} = 2 \pounds_n \sigma_{ab} - \rho_{\la ab\ra}$; $n$ is the null tangent to $\scri$, and $\rho_{\la ab\ra}$ the trace-less part of Geroch's tensor, which vanishes if the conformal frame is a round sphere, also known as a Bondi frame. 
The explicit form of the BMS transformation $\d_\xi \s_{ab}$ can be found in the references, and will not be needed.
The foliation induces a family of Lorentz subgroups of the BMS group, one per cross-section. Picking 
an auxiliary 1-form $l_a$ transverse to $n$ and adapted to the leaves of the affine foliation, we can parametrize $\xi\eqonS f n+Y$, where $l\cdot Y=\pounds_n Y=0$ and $\pounds_n f=\f12\Dd_a Y^a$. Here $Y^a$ is the Lorentz parameter, represented by a globally defined conformal Killing vector on the cross-sections, and the time-independent part of $f$, let us denote it $T$, is the super-translation parameter. The cross-sections have the sphere topology, with time-independent metric $q_{ab}$, and we denote $\Dd_a$ its covariant derivative and $\cR$ its Ricci scalar.
We can then write the charges on any cross section with an adapted $l$ as
\begin{align}\label{qBMS}
Q^{\sscr BMS}_\xi = \f1{8\pi G}\oint_S (2f  M_\r+ Y^a J_a)\eps_S, \qquad M_\r:=M+\f14\r_{ab}\s^{ab}.
\end{align}
They correspond to Geroch's super-momentum \cite{Geroch:1977jn} and to the Dray-Streubel Lorentz charges \cite{Dray:1984rfa}, as proved in 
\cite{Ashtekar:1981bq,Dray:1984gz}. The aspects $M$ and $J_a$ can be written using a Newman-Penrose tetrad at $\scri$, see e.g. \cite{Dray:1984rfa,Rignon-Bret:2024gcx}, as
\be\label{MNP}
M = -\left( \psi_2+\s\dot{\bar\s} +\f12\left(\eth^2 \bar\s - cc\right)\right) = -\re(\psi_2 + \s \dot{\bar\s}), 
\ee
and 
\be\label{DS}
m^a J_a = -\left( \psi_1 + \s\eth \bar\s+\f12\eth(\s\bar\s)\right).
\ee
Here $\s:=-m^am^a\s_{ab}$ and we are using conventions adapted to a mostly-plus metric signature, and the shear and Weyl scalars refer to their leading-order asymptotics, but we omitted the apex $^0$ typically used to label them for ease of notation.
It is also possible to write them in terms of tensors, see e.g. \cite{Geroch:1977jn,Ashtekar:2024bpi}, or in terms of the metric components of an asymptotic expansion in Bondi coordinates, see e.g. \cite{Barnich:2011mi,Flanagan:2015pxa,Rignon-Bret:2024gcx}. 
The link to Bondi coordinates can be made explicit choosing coordinates $x^a=(u,x^A)$ with $u$ an affine parameter so that $n\eqonS\p_u$ and $l=-du$, then all tensors appearing in \eqref{qBMS} have vanishing $u$ components, and $f=T+\f u2\Dd_A Y^A$.

The original papers singled out the unique choices \eqref{ASflux} and \eqref{qBMS} by a series of physical requirements: they are conserved in non-radiative spacetimes; they vanish in Minkowski spacetime, for any $\xi$ and at any cut; they are covariant, namely independent of coordinate choices, as well as of other background structures such as the choice of conformal factor and of foliation of $\scri$.\footnote{The only non-dynamical quantities they can depend on are the cross section $S$ and the symmetry vector field $\xi$.} There were also technical requirements, such as depending on a number of derivatives compatible with second order field equations, and be related to a choice of canonical symplectic form associated with the Einstein-Hilbert Lagrangian. The later Wald-Zoupas insight \cite{Wald:1999wa} is that they are Noether charges for a specific choice of symplectic potential at $\scri$, given by
\be\label{thBMS}
\th^{\sscr BMS} = -\f1{16\pi} N_{ab}\d \s^{ab}\eps_\scri,
\ee
so that the integrand of the flux \eqref{ASflux} is the Noether current $j_\xi=I_\xi\th^{\sscr BMS}$,
and that they are related in a precise sense to symmetry generators on a space-like hypersurface $\Si$ intersecting $\scri$ at $S$.
Finally, they provide a realization of the symmetry algebra in terms of the Barnich-Troessaert  bracket \cite{Barnich:2011mi} as
\begin{align} \label{Qcov}
\{Q_\xi, Q_\chi\}_*  :=\d_\chi  Q_\xi - \oint i_\chi j_\xi 
\eqons Q_{[ \xi,\chi]}. 
\end{align}
This property can be explicitly checked \cite{Rignon-Bret:2024gcx}, and shown to be a consequence of the covariance of \eqref{thBMS} and of the BMS boundary conditions \cite{Rignon-Bret:2024wlu}. It generalizes the standard Hamiltonian action of a symmetry to dissipative situations with a non-zero current at the cross section, and reduces to it at non-radiative cuts.

Let us now focus attention on the super-translation charges. Their flux is 
\be\label{QTflux}
Q^{\sscr BMS}_T[S_2]-Q^{\sscr BMS}_T[S_1] \eqons - \f1{32\pi}\int\left(TN^{ab}N_{ab} + 2N^{ab}(\Dd_a\Dd_b+\f12\r_{ab})T  \right) \eps_\scri.
\ee
The first is the `hard term', squared in the news. The second is the `soft term', linear in the news. It follows from the time-independence of $T$ and of the volume form that  the soft term corresponds to the displacement memory \cite{Christodoulou:1991cr}
\be\label{memoryterm}
{\mathbbm{ M}} 
=-\f1{8\pi}\oint T{\mathbbm{ m}}\eps_S, \qquad {\mathbbm{ m}} := \DDr^{ab}({\s}_{ab}-\f u2\r_{ab})\big|^{S_2}_{S_1}.
\ee 
Here we introduced the short-hand notation $\DDr{}_{ab}:=\Dd_{\la a}\Dd_{b\ra}+\f12\r_{\la ab\ra}$. When acting on $T$, the four zero modes of this operator define global translations, which form an ideal of the BMS algebra, and can be identified 
with the first four harmonics $l=(0,1)$ in Bondi frames. It follows that the soft term vanishes for global translations, resulting in a `purely hard' flux. Furthermore, it is strictly negative for a future-pointing global time translation: 
this is Bondi's celebrated result proving that gravitational waves dissipate energy. 

Modifying the flux-balance law so that it has a purely hard flux for \emph{any} $T$ is at first sight straightforward.
Since the soft term is a total derivative in $u$, it is possible to reabsorb it in the left-hand side of \eqref{QTflux}, and consider the charge
\be \label{QM}
Q^{\sscr M}_T = \f1{4\pi}\oint_S T\Big(M_\r + \f12 {\DDr}^{ab} (\s_{ab}-\f u2\r_{\la ab\ra})\Big) \eps_S.
\ee
If we restrict this expression to a Bondi frame, we recognize
the Moreschi mass
\be \label{Moreschimass}
	M^{\sscr M} := M + \f12 {\Dd}_a {\Dd}_b \sigma^{ab},
\ee
that was proposed as a super-momentum charge in  \cite{Moreschi:1998mw,Dain:2000lij}.
 The problem with this proposal is that it is manifestly non-covariant. Adding Geroch's tensor fixes the limitation of the original expression to Bondi frames and achieves full conformal invariance. However it is still foliation dependent, or equivalently $l$-dependent, via the shear and the affine parameter $u$. Foliation-dependence carries an intuitive meaning of non-covariance, which can be made sharper observing that two super-translation vector fields commute, hence a foliation-dependent super-momentum charge would fail to capture this basic property of the BMS algebra. The lack of covariance of \eqref{Moreschimass} can be made explicit computing its transformation law. We find
  \be
  \d_\chi Q^{\sscr M}_T \eqons  Q^{\sscr M}_{[\xi_T,\chi]} + K_{(\xi_T,\chi)} + {\rm radiation \ terms}, 
  \ee
where 
\be
K_{(\xi_T,\chi)} = \f1{16\pi} \oint_S T\left( (\Dd^2+\cR)\Dd^2 f + 2 \Dd_a\cR\Dd^a f - 4\s^{ab}\Dd_a\Dd_b f\right)\eps_S
\ee  
for an arbitrary BMS generator $\chi = f n + Y$. 
This term prevents the recovery of the correct symmetry action on non-radiative cuts, and gives
a field-dependent cocycle on the right-hand side of the Barnich-Troessaert bracket.
In the next two sections we show how this problem can be resolved, and a covariant charge with a purely hard flux obtained. The first step is to use the  `covariant shear' of \cite{Compere:2018ylh}.

\section{Super-translation Goldstone and covariant shear}

The AS radiative phase space is supplemented with the late time boundary conditions
\be\label{bc}
\lim_{u\to\infty}	N_{ab} \sim \frac{1}{u^{1 + \eps}}, \qquad \lim_{u\to\infty}\im(\psi_2) \sim \f 1{u^{\eps}},  \qquad \eps > 0,
\ee
namely no radiation and a vacuum shear $\sv$ that is purely electric,
\be\label{sfinal}
\lim_{u\to\infty}	\s_{ab} - \f12 u \rho_{\la ab\ra} =\sv_{ab} - \f12 u \rho_{\la ab\ra}= -({\Dd}_{\la a} {\Dd}_{b\ra} + \frac{1}{2} \rho_{\la ab\ra}) u_0.
\ee
The system thus settles down to equilibrium, and the radiative degrees of freedom reduce to a vacuum parametrized by $u_0(x^A)$, whose four-parameter family of zero modes are the famous `good cuts'. The late-time boundary datum $u_0$ for the shear can be interpreted as a choice of `bad cut' \cite{Rignon-Bret:2024gcx}, as a super-translation field parametrizing the vacua \cite{Compere:2018ylh}, or as Goldstone mode for the breaking of the super-translation symmetry \cite{Strominger:2013jfa} caused by the choice of final vacuum state $\sv$. 
Its transformation rule, induced by that of a vacuum shear, is \cite{Rignon-Bret:2024gcx}
\be \label{dxiu0}
\d_\xi u_0 =  \pounds_\xi u_0 - T - u_0 \dot{f} = Y^a \p_a u_0 - T - u_0 \dot{f}.
\ee
We can use the boundary condition \eqref{sfinal} to define the relative shear 
\be\label{defS}
\cS_{ab} := \sigma_{ab} - \sv_{ab} = \sigma_{ab} - \f12 u \rho_{\la ab\ra} + ({\Dd}_{\la a} {\Dd}_{b\ra} + \frac{1}{2} \rho_{\la ab\ra}) u_0 \equiv -\f12{\cal C}_{ab},
\ee
which is nothing but the `covariant shear' ${\cal C}_{ab}$ introduced in \cite{Compere:2018ylh}. The term covariant refers to the fact that it is invariant under super-translations, hence independent of the background structure given by the $u$ foliation. This property follows from the definition, and can be verified using \eqref{dxiu0}. 
The boundary condition \eqref{sfinal} implies that  the covariant shears vanishes at late times,
\be \label{asymptoticrelshear}
	\lim_{u\to\infty}	 \cS_{ab} = 0.
\ee
It furthermore vanishes exactly in any non-radiative spacetime, since the shear has to match its own `bad cut' in the absence of radiation. 
The relative shear has conformal weight 1, and satisfies 
\be\label{NewsS}
N_{ab} = 2 \pounds_n \cS_{ab}.
\ee
It is thus a covariant choice of `time potential' for the news, and can be used to write the memory \eqref{memoryterm} as
\be
{\mathbbm{ m}} = \DDr^{ab}{\cS}_{ab}\big|^{S_2}_{S_1}.
\ee

It is  natural to ask whether one can use the ambiguity to add time-independent terms to the charges to replace $\s_{ab}$ in \eqref{QM} with ${\cal S}_{ab}$ and achieve covariance in this way. This cannot be done in the context of \cite{Ashtekar:1981bq,Wald:1999wa}, since already the inclusion of the (double derivative of the) shear in \eqref{QM} has spoiled the relation to the canonical choice for the symplectic 2-form. The BMS charges are unique, after all \cite{Ashtekar:2024bpi}. This leads us to the second step, described next.

\section{Corner improvement and new charges}

The way out of the  uniqueness result for the standard BMS charges is that we will allow for corner modifications of the symplectic 2-forms. These modifications are compatible with the field equations, and can be motivated by situations in which stationarity and covariance, or finiteness, cannot be achieved otherwise, see e.g. \cite{Compere:2008us,Harlow:2019yfa,Freidel:2020xyx,Campiglia:2020qvc,Odak:2021axr,Chandrasekaran:2021vyu,Odak:2022ndm,Donnay:2022hkf,Rignon-Bret:2024wlu}.
For BMS all these properties are already satisfied with the standard symplectic 2-form, and our reason to consider a different one comes from the additional requirement of having a purely hard flux.

The AS symplectic 2-form is
\be\label{omAS}
\Om^{\sscr AS} = -\f1{16\pi}\int \d N_{ab}\curlywedge \d \s^{ab}\eps_\scri,
\ee
and corresponds to the value at $\scri$ of the standard Einstein-Hilbert symplectic 2-form $\Om^{\sscr EH}$ given for instance in \cite{Wald:1999wa}.
The integral here can be over a portion of $\scri$ or all of it if we include the boundary conditions given in the first equation of \eqref{bc}.
The new symplectic structure that we propose is
\be\label{barOm}
\bar \Om_\scri = -\f1{16\pi}\int \d N_{ab}\d \cS^{ab}\eps_\scri = \Om^{\sscr AS} - \d\int d\vth, 
\ee
where 
\be\label{vthdef}
\vth = \f1{8\pi}\cS_{ab}\d(\DDr^{ab} u_0)\eps_S = \f1{8\pi}\cS_{ab}\, \d\sv^{ab}\eps_S.
\ee
The corner term depends explicitly on the electric vacuum late-time boundary condition. 
We observe that it
makes the Goldstone mode $u_0$ canonically conjugated to the displacement memory effect, since
\be
\d\int_{S_1}^{S_2}d\vth = \f1{8\pi}\oint \d {\mathbbm{ M}}\d u_0 \eps_S, 
\qquad {\mathbbm{ M}} := \DDr^{ab}{\s}_{ab}\big|^{S_2}_{S_1}.
\ee
To be precise, it is only the super-translation modes that are canonically conjugated to the memory, because the zero modes drop out of the integral, in agreement with the fact that the corner term is a function of the vacuum shear and not of a specific choice of $u_0$.\footnote{The corner symplectic structure involves the same symplectic pair as the bracket posited in \cite{He:2014laa}, and it would be interesting to investigate further if there is a relation with our symplectic structure.}

To write the fluxes and charges we need on top of the symplectic 2-form also a choice of preferred symplectic potential.\footnote{For BMS, this step can be replaced by a choice of topology in the phase space \cite{Ashtekar:2024bpi}. It would be interesting to know if this alternative exists also for the new symplectic structure.} We take
\be\label{thnew}
\bar\th=\th^{\sscr BMS}-d\vth= -\f1{16\pi}N_{ab}\d\cS^{ab}\eps_\scri,
\ee
where in the second equality we used \eqref{NewsS}.
This choice is manifestly covariant and vanishes in non-radiative spacetimes, hence it satisfies the conditions for the generalized Wald-Zoupas prescription
\cite{Wald:1999wa,Rignon-Bret:2024wlu}.
The associated fluxes are
\be
\bar F_\xi = -\f1{16\pi}\int N_{ab}\d_\xi \cS^{ab}\eps_\scri = F^{\sscr BMS}_\xi - \f1{8\pi}\oint^{S_2}_{S_1}  \d_\xi u_0\, \DDr^{ab}{\cal S}_{ab}\, \eps_S,
\ee
with charges at constant $u$ cross-sections
\be\label{barQ}
\bar Q_\xi = \f1{8\pi}\oint_S [(2f M_\r + f|_{u_0} {\DDr}^{ab}\cS_{ab})+  Y^a (J_a - \DDr^{bc}\cS_{bc}\, \Dd_a u_0)]\eps_S.
\ee
Here $ f|_{u_0} = T + u_0 \dot{f}$. Since the symplectic potential is Wald-Zoupas, the fluxes are guaranteed to be covariant, and the charge algebra admits at most a time-independent cocycle \cite{Rignon-Bret:2024wlu}. Furthermore the original BMS algebra is center-less \cite{Rignon-Bret:2024wlu,Rignon-Bret:2024gcx}, hence the only cocycle can come from an anomalous $\vth$ in the shift \cite{Chandrasekaran:2020wwn,Freidel:2021cjp,Freidel:2021yqe,Chandrasekaran:2021vyu}. But the choice \eqref{vthdef} is covariant, as it can be easily verified, hence the new charges \eqref{barQ} are also covariant, and satisfy \eqref{Qcov} without any cocycle or central extension.

The new charges differ from the standard BMS charges only in radiative spacetimes, where $\cS_{ab}\neq 0$. Hence they automatically have the properties of being conserved in non-radiative spacetimes, and to vanish exactly in Minkowski spacetime, for any symmetry parameter, at any cut. They therefore satisfy the same requirements of the standard BMS charges. What distinguishes them is only the different symplectic structure used.
Notice that the modified symplectic 2-form can  be used also in the case of a space-like hypersurface $\Si$ intersecting $\scri$ at the cross-section $S$,
\be
\bar\Om_\Si := \Om^{\sscr EH}_\Si -\d \oint_S \vth.
\ee
Therefore the new charges can  be considered Hamiltonian generators in the same sense as in the original Wald-Zoupas construction.
 
The new super-momentum charge is
\be\label{barQT}
	\bar Q_T = \f1{4\pi}\oint_S T\bar M\eps_S, \qquad \bar M:= M_\r + \f12 \DDr^{ab}{\cal S}_{ab}.
\ee
It is covariant, a canonical generator of super-translations with respect to the corner-improved symplectic structure \eqref{barOm}, and has a purely hard flux
\be\label{QTfluxnew}
\bar Q_T[S_2]-\bar Q_T[S_1] \eqons - \f1{32\pi}\int TN^{ab}N_{ab} \eps_\scri.
\ee
It can be seen as 
a version of the Moreschi mass made
covariant under conformal transformations and super-translations thanks to $\r_{ab}$ and $u_0$. The precise relation between the two is 
\be
\bar M = M^{\sscr M} + \f1{8\pi}\oint_S T\left( \r^{ab}{\s}_{ab} -\f u2\DDr^{ab}\r_{ab}+\DDr^2 u_0\right)\eps_S.
\ee
On round spheres, the extra term achieving foliation-independence reduces to $\oint T (\Dd^2+2) \Dd^2 u_0$, and vanishes for global translations.

There are two differences between our proposal and the one of \cite{Moreschi:1998mw,Dain:2000lij}, a minor one and a major one. The minor
one is that the Moreschi mass $M^{\sscr M}$ has a purely hard flux only when the asymptotic frame is a round sphere, aka Bondi frame. This is fixed adding the contribution of Geroch's tensor (and we can remark the linear time-dependence that it brings in), so that the flux is purely hard in any (divergence-free) frame. Since for BMS boundary conditions we can always choose a Bondi frame, this is just a minor improvement. 
A major difference is that $M^{\sscr M}$ is foliation dependent, and for this reason it fails to satisfy the Wald-Zoupas
covariance requirement and does not provide a charge satisfying the BMS algebra in the phase space. From the perspective of \cite{Moreschi:1998mw,Dain:2000lij}, this is not a problem but instead a desired feature, since the goal there was precisely a foliation-dependent quantity that could be used for their line of work on so-called `nice cuts'. It is a feature that makes $M^{\sscr M}$ useful for instance in numerical relativity  \cite{Mitman:2021xkq}.
And the fact that $M^{\sscr M}$ has a purely hard flux (on Bondi frames) was an observation and not a
goal of the construction. Conversely, we are interested in the
opposite problem of having a foliation-independent quantity (hence
Wald-Zoupas covariant and providing a realization of the algebra),
while preserving the purely hard flux. Since the flux is the same, the
only difference can be a time-independent term; our key result is that
there exists indeed a time independent term that removes the
foliation-dependence from $M^{\sscr M}$.
As a consequence of the additional time independent term, our formula
does \emph{not} reproduce the one of \cite{Moreschi:1998mw,Dain:2000lij} even in Bondi frames. 
For instance, the Moreschi mass does not vanish in Minkowski spacetime if we choose a bad cut, since the shear does not vanish there, while our charge vanishes in Mikowski at any cut, a covariance property that it shares with Geroch's super-momentum.
At the level of the algebra, the difference is manifest in the charge algebra formula
$\d_{T'} \bar Q_{T} =0$ for non-radiative spacetimes and for any two super-translations, which is satisfied by our
proposal (and by Geroch's super-momentum) but not by the one of  \cite{Moreschi:1998mw,Dain:2000lij}, nor by any other foliation-dependent quantity.\footnote{Of course, given a solution, one can always find a foliation in which the two proposals match (on round spheres). But this is like saying that a gauge-dependent quantity can always be made to match its gauge-independent part with a suitable choice of gauge. 
Our point is that there exists a formula that has the same value no matter which foliation is chosen, and only this quantity will realize the charge algebra.}
Being foliation-independent, \eqref{barQT} cannot on the other hand
be used to fix a `BMS frame' in the sense of \cite{Mitman:2021xkq}, namely a
reference foliation. For the goals of that reference, the relevance of the Moreschi mass is untouched, in our opinion.

The new charges are also unique, in the sense that they are the only ones with the purely hard flux satisfying the covariance and stationarity requirements. 
To see this, we notice first that for a given flux, the only potential ambiguity in the charges is the addition of time-independent terms to $\vth$. 
But the only time independent fields are $u_0$, $\r_{ab}$ and $q_{ab}$, and a moment of reflection shows that it is not possible to write something conformally invariant and foliation independent using only these fields. 
Hence any ambiguity is ruled out by the requirement to have covariant charges.
Second, similar considerations show that \eqref{thnew} is the only symplectic potential with a purely hard flux which satisfies the Wald-Zoupas requirements of covariance and stationarity, and respects the order of derivatives of the field equations. We conclude that the symplectic potential \eqref{thnew} and charges \eqref{barQ} are unique. What distinguishes them from their standard BMS analogue is only the modification of the symplectic structure.

Let us add a remark about the boundary conditions. If memory is present, the relative shear can only vanish in one of the two limits $u\to\pm\infty$. 
This means that the new charges match the standard BMS charges also in only one of the two limits.
Our construction works with both options, relative shear vanishing at early or late time. The interest in choosing the late time option \eqref{asymptoticrelshear} is to reproduce the standard BMS charges at the end of the evolution, e.g. the Kerr charges for a typical coalescing binary system, and vanishing charges if the end state is empty with all the energy having radiated away to $\scri$ as in scattering spacetimes. 
It is also useful in order to prove that the charges associated with future pointing super-translations positive on any cut, as we will show below.

\section{Spatially local energy}

We now discuss the main point of our proposal, namely the fact that it leads to a new characterization of (spatially) local gravitational energy.
With the standard BMS charges, only the total energy has a monotonic flux. This corresponds to the lowest mode of the super-translation parameter, or $T=1$ on a Bondi frame. Consider instead a localized BMS super-momentum charge, corresponding for instance to a parameter $T$ in \eqref{qBMS} with a Gaussian-like profile peaked on some point on the celestial sphere. This profile necessarily involves higher modes -- whose flux is not monotonic, hence the charge cannot be interpreted as a localized energy, because it may be increasing even though the system is dissipating. Indeed, the standard interpretation of the  BMS super-momentum higher modes is not related to energy, but rather to the notion of mass multipoles. Their flux is not monotonic even for positive $T$ and captures the memory contribution.

The new charge \eqref{barQT} has instead a flux which is monotonic for any positive super-translation parameter, and independent of memory. 
It follows that we can take a peaked Gaussian-like function $T$ localised as sharply as we want on the celestial sphere, and the associated charge will necessarily decrease in any dynamical process! This allows us to extend Bondi's result about dissipation from the total energy to a spatially local energy density,
given by \eqref{barQT} with a sharply peaked $T$, e.g. a Dirac delta $T_\circ=\d^{(2)}(x,x_\circ)$.
The result is a charge
\be\label{localcharge}
\bar Q_{T_\circ}(u,x_\circ) = \f1{4\pi}\oint_S T_\circ\bar M\eps_S = \f1{4\pi}\bar M|_{x_\circ},
\ee
localized on the celestial sphere and with a monotonic flux localized on a single null geodesic,
\be\label{localflux}
\bar Q_{T_\circ}(u_2,x_\circ)-\bar Q_{T_\circ}(u_1,x_\circ) \eqons - \f1{32\pi}\int_{u_1}^{u_2}du\left(N^{ab}N_{ab}  \eps_S \right)|_{x_\circ}.
\ee

The crucial question is whether it is possible to interpret it as an energy. To support this proposal, we point out that it meets various criteria that one would expect from energy.
First, it is the Noether charge and canonical generator for the associated super-translation symmetry, with an admissible choice of symplectic structure. Second, it is always positive, thanks to the negativity of the flux, if we assume that the final state has a positive Bondi mass aspect, as is the case for a stationary black hole.\footnote{This assumptions is guaranteed if we strengthen the boundary conditions replacing 
 $\lim_{u\to\infty}\im(\psi_2)=0$ with $\lim_{u\to\infty}\dot\psi_1=0$ in the rest frame \cite{Ashtekar:2019viz}. The latter condition in fact implies not only the former, but also that $\psi_2$ is spherically symmetric, hence positive since the total Bondi energy is positive. Boosting the frame loses the spherical symmetry but not the positivity since the transformation is homogeneous in the non-radiative region and the boost parameter is positive. Hence positivity holds in any frame. }
 Third, it is extensive. This follows from the fact that it is functionally linear in $T$, hence the total energy in two small regions identified by two sharply-peaked parameters is the charge whose parameter is the sum of the two parameters. 
Extensivity may come as a surprise, since the theory is non-linear and it is not possible to screen gravitational interaction. But here we are talking about extensivity at $\scri$, where different regions on the celestial sphere are infinitely far away and \emph{causally disconnected}.
This is what makes extensivity possible, and indeed natural.\footnote{A similar construction could be done also on local null hypersurfaces if one uses a monotonic flux, however only within domains free of caustics or crossings, as to guarantee causal disconnection. Otherwise extensivity would no longer be possible. Super-translations are in fact not even admissible symmetries of local null surfaces \cite{Chandrasekaran:2018aop}.}  

It should be stressed that our proposal is perfectly compatible with the equivalence principle, which implies that a complete localization of energy is not possible. First, because we only used properties of general relativity without any modification, and second because indeed, some aspects of non-locality do remain. Specifically, the localization obtained for the proposed energy is only spatial and not temporal, hence not complete. Let us discuss this subtle point, which is central to our proposal, in more detail.

Thanks to the structures present at $\scri$, we have been able to construct the spatially local quantity \eqref{localcharge} and shown that it has properties that support its interpretation as an energy. And furthermore its flux \eqref{localflux} is also a local function of the news tensor. But the catch is that the news tensor is a \emph{non-local} function of the Weyl curvature, in both space and time. It is indeed the {potential} of (some components of) the Weyl curvature at $\scri$. This can be conveniently formulated using for example the Newman-Penrose formalism,
\be\label{psi34}
\psi_3  
=\eth N, 
\qquad \psi_4
=\dot N,
\ee
where $N:=\f12\bar m^a\bar m^b N_{ab}$ and $N_{ab}N^{ab}=4N\bar N$. Therefore measuring the flux of the local energy still requires non-local knowledge of the curvature. 
  For this reason, our construction does not achieve a complete localization, in agreement with the principle of general covariance. It is nonetheless 
  remarkable that the non-locality can be limited to \eqref{psi34}. 
In particular, since the news can be determined using only the second equation in \eqref{psi34},  it is possible to determine \eqref{localcharge} doing measurements that are purely local in space, and non-local only in time:
\be \label{newsint}
N(u,x_\circ) = - \int_u^{+ \infty} \psi_4(u',x_\circ) du'.
\ee
Using this equation on the right-hand side of \eqref{localflux}, one can determine the variation of the local energy doing only spatially local measurements along the geodesic at $x_\circ$ over the extended period of time until the source stops radiating.\footnote{The non-radiating late-time boundary condition is used explicitly in \eqref{newsint}. One may worry that this spoil spatial locality, since verifying that the news vanishes requires a non-local measurement in space. However our perspective is that vanishing news at late time is an assumption used in defining the phase space, and not a quantity to be observed or measured.}
As for the local energy at a given instant of time, it can be determined in terms of this variation plus an initial datum. Taking the datum at late times where the stationarity boundary conditions hold, it is entirely determined by the local measurement of $\re(\psi_2)$. 
Therefore the candidate energy is truly spatially local.

\section{Conclusions}

The new BMS charges presented here satisfy the same physical requirements as the standard ones of \cite{Geroch:1977jn,Ashtekar:1981hw,Ashtekar:1981bq,Dray:1984rfa,Dray:1984gz}:
covariance (namely independence from coordinates and background structures), being conserved in any non-radiative spacetimes, and vanishing in Minkowski for any symmetry generator and at any cut. What distinguishes them is only a different choice of symplectic structure. Both choices are compatible with the same field equations. 
Our perspective is that both choices are mathematically and physically viable, and picking one over the other is akin to 
taking different boundary conditions or polarizations of the phase space: it depends on the specific problem one is interested in.
Indeed if the charges are observables in general relativity, their relevance should be related to an experiment and to the physical quantity that one is measuring. 
From this perspective, the standard BMS charges are relevant to study the memory effect, and the new ones to study the energy dissipated in an arbitrarily small region of the celestial sphere.
For other examples of the dependence of charges on the boundary conditions in gravitational systems, see e.g. \cite{brown1990thermo,Freidel:2021cjp,Chandrasekaran:2021vyu,Adami:2021kvx,Odak:2021axr,Odak:2023pga,Chandrasekaran:2023vzb,Rignon-Bret:2024zhj}.

Let us mention two potential further applications of a monotonic flux for all future-pointing super-translations.
The first is related to the work of Wall \cite{Wall:2011hj}.
His proof of the generalized second law requires the axioms of 
ultra-locality and stability. Ultra-locality means that the null geodesics spanning a null hypersurface can be treated as independent subsystems, and stability implies that the Hamiltonian on each subsystem has positive eigenvalues. Identifying the Hamiltonian density with the super-translation Noether current $j_T$, we can state the axioms as 
\be\label{posflux}
	\int_{\mathcal{N}} j_T  \eps_\mathcal{N} \geq 0,
\ee
for any $T>0$, which includes any approximation of a delta function localized at a point.
In Wall's work the null hypersurface $\cN$ is an horizon, for which a monotonic flux has been studied in e.g. \cite{Ashtekar:2021kqj,Rignon-Bret:2023fjq,Odak:2023pga,Ciambelli:2023mir,Hollands:2024vbe}.
Our work shows that the axioms can apply to $\scri$ as well, if the right symplectic structure is chosen. This is in line with the interpretation of $\scri$ as an horizon (in the conformal spacetime \cite{Ashtekar:2024mme,Ashtekar:2024bpi}), and can be relevant for studying the generalized second law at $\scri$ (e.g. \cite{Bousso:2016vlt,Faulkner:2024gst}).

A second potential application comes from soft theorems, where it was pointed out that the soft terms in the flux are related to infra-red divergences \cite{Ashtekar:2018lor,Arkani-Hamed:2020gyp}. 
This motivated the work of \cite{Campiglia:2021bap,Donnay:2022hkf} where the hard contribution in the flux was isolated from the soft one already at the level of the symplectic potential. The result was generalized in \cite{Rignon-Bret:2024gcx} providing a current for this symplectic potential and establishing that it satisfies the Wald-Zoupas criteria. It is given by
\be\label{theBMS}
\th^{\sscr eBMS} := \th^{\sscr BMS} -d \vth^{\sscr eBMS}
= - \frac{1}{16 \pi} \left(N_{ab} \d \s^{ab} +\f12 (u - u_0) N_{ab} \d \rho^{ab}\right) \eps_\scri.
\ee
Using \eqref{defS} and the standard eBMS boundary condition $\d q_{ab}=0$, we can rewrite it as
\be\label{theBMS1}
\th^{\sscr eBMS}
= -\frac{1}{16 \pi} (N_{ab} \d {\cal S}^{ab} + (u - u_0) N_{ab} \d \rho^{ab} -  \d u_0\DDr{}_{ab}N^{ab}) \eps_\scri.
\ee
Comparing \eqref{theBMS1} to \eqref{thBMS}, the isolation of the hard term (quadratic in the radiative fields and super-translation invariant) from the soft term (linear in the radiative fields and super-translation dependent) is manifest. 
There are however two important differences between this line of research and the results presented instead in the current paper. First, what we have achieved here is not simply a split hard-soft at the symplectic level as in \eqref{theBMS1} and \cite{Campiglia:2021bap,Donnay:2022hkf}, but rather a \emph{removal} of the soft term from the flux using the corner improvement \eqref{vthdef}. Second, what we have done is only possible for the BMS boundary conditions, and not for the eBMS ones. The key difference is that the symmetry vector fields singularities allowed in the latter case render Geroch's tensor not universal, hence the second term in \eqref{theBMS} and \eqref{theBMS1}. While we can remove the memory term in  \eqref{theBMS1} using the same corner improvement \eqref{vthdef}, it is not possible to remove the term with Geroch's tensors using the freedom to change the symplectic potential,\footnote{One can formally write the second term as a total time derivative introducing a primitive for $uN$, however the resulting corner improvement would be a function of the fields everywhere and not just at the corner. } and this is still soft and super-translation dependent.
Hence a purely hard flux can only be achieved with BMS boundary conditions and not eBMS ones. 

As a concluding remark, we should stress the importance of being at null infinity in order to be able to construct a well-defined, spatially local energy density of the gravitational field. Spatially localised observers on the celestial sphere are causally disconnected, therefore it is natural for each of them to construct  their own generator of time translations. Our results show that it can be done in an unambiguous way.

\providecommand{\href}[2]{#2}\begingroup\raggedright\endgroup


\end{document}